\documentclass[prl,twocolumn,longbibliography,letterpaper,showpacs,groupedaddress,superscriptaddress,nofootinbib,floatfix,preprintnumbers,tightenlines]{revtex4-1}
\usepackage{amsmath,amssymb,amsfonts}
\usepackage{graphicx}
\usepackage{bm}
\usepackage{hyperref}
\usepackage{comment}
\usepackage{color}
\usepackage{csquotes}
\usepackage{longtable}
\usepackage[section]{placeins}

\allowdisplaybreaks
\let\Oldsection\section
\renewcommand{\section}{\FloatBarrier\Oldsection}
\let\Oldsubsection\subsection
\renewcommand{\subsection}{\FloatBarrier\Oldsubsection}

\newcount\hour \newcount\hourminute \newcount\minute
\hour=\time \divide \hour by 60
\hourminute=\hour \multiply \hourminute by 60
\minute=\time \advance \minute by -\hourminute
\newcommand{\mydate}{\ \today \ - \number\hour :\number\minute}
\newcommand{\bea}{\begin{eqnarray}}
\newcommand{\eea}{\end{eqnarray}}

\newcommand{\eq}[1]{Eq.~(\ref{#1})}
 
\newcommand {\bfsigma}{\mbox{\boldmath$\sigma$}}

\def\r{\rho}\def\a{\alpha}\def\d{\delta}\def\th{\theta}

\def\OMIT#1{{}}

\def\D{\Delta}

\def\r{\rho}
\def\ra{\rangle}\def\la{\langle}

\def\si{^1 \hskip -0.03in S _0}
\def\up{\uparrow}\def\down{\downarrow}
\def\s{\sigma}
\def\siii{^3 \hskip -0.025in S _1}

\def\diii{^3 \hskip -0.03in D _1}
\def\bfp{{\bf p}}\def\bfn{{\bf n}}
\def\bfK{{\bf K}}\def\bfp{{\bf p}}\def\O{\Omega}\def\ph{\phi}
\def\t{\tau}

\begin{document}

\title{
Entanglement Maximization  in Low-Energy  Neutron-Proton Scattering}

\preprint{NT@UW-23-05}

\author{Gerald A.~Miller}
\affiliation{Department of Physics, University of Washington, Seattle, WA 98195-1560, USA}

\date{\mydate}

\begin{abstract}
  \noindent
 The entanglement properties of neutron-proton scattering are investigated using  a measure that counts the number of entangled pairs produced by the action of a scattering operator on a given initial neutron-proton state. All phase shifts relevant for scattering at laboratory energies up to 350 MeV are used.
 Entanglement is found to be maximized in very low energy scattering. At such energies the Hamiltonian obeys Wigner SU(4) symmetry, and an entanglement maximum is a sign of that symmetry. At higher energies the angular dependence of entanglement is strong and the entanglement is large for many scattering angles. The tensor force is shown to play a significant role in producing entanglement at lab kinetic energies greater than about 50 MeV.

  \end{abstract}
\maketitle

 Stimulated by the connection with quantum computing, which rests on the possibility that entanglement may enhance computing capabilities, 
the implications of entanglement in quantum mechanics and quantum field theory have recently been studied in many papers.
For a long list of recent references see Ref.~\cite {Ehlers:2022oal}. Ideas  related to  quantum entanglement provide a new way of looking at old problems and may provide insights into deep connections with underlying symmetries.   For example, 
 Refs.~\cite{Cervera-Lierta:2017tdt,Fedida:2022izl}  argued that a principle of {\it maximum} entanglement  is responsible for the particular sets of coefficients that define quantum electrodynamics. Similarly Refs.~\cite{Kharzeev:2017qzs,PhysRevC.99.015205,PhysRevD.102.074008,Hentschinski:2022rsa}
 argue that  high-energy interactions involve  maximally entangled states.
 Maximum entanglement  is a property of  nucleon valence quark distributions~\cite{PhysRevD.91.054026}, and large entanglement entropy  is a property of the nucleon state vector~\cite{Beane:2019loz}.
 On the other hand Ref.~\cite{Beane:2018oxh} (BKKS)
proposed that nucleon-nucleon scattering is described by entanglement suppression that is correlated with Wigner SU(4) symmetry~\cite{Wigner:1936dx}.  See also~\cite{Bai:2023tey}.  Wigner used this  describe the low-lying spectra of light nuclei. Similar statements regarding entanglement suppression appear in~\cite{Liu:2022grf}.

The purpose here is to  provide a more detailed  study of entanglement entropy  in neutron-proton scattering.  Let's begin with some basic issues.
The textbook~\cite{Gottfried} definition of  
entropy, the von Neumann entropy, given by
$ S=-{\rm Tr}[\r \log\r],$ where $\r$ is the   density matrix. The operator $\r$  can be diagonalized, with eigenvalues   designated as $p_n$ and $\sum_n p_n=1$.  In this diagonal representation   $S$ is expressed as:
\bea S=-\sum_{n=1}^d p_n \log p_n,  
\label{SvN}\eea
where $d$ is the dimension of the space. 

The quantity $S$ is maximized when all of the probabilities are equal: $p_n=1/d$.
In that case $ S_{\rm max} =\log d.$   The value of $d=2$ for a particle of spin 1/2.  This situation of maximum entropy is one of no entanglement. If all of the probability eigenvalues are the same, the density matrix is given by 
$\r_{\rm max}={\hat I\over d} $ where  $\hat I$ is the identity operator.  This is known as the classical or ``garbage state"~\cite{Bennett:1996gf}. 

Instead the amount of entanglement of a state $, |\ph\ra$, of two spin=1/2 particles  is measured by computing the amount of overlap with completely entangled Bell states:
\bea |e_1\ra={1\over \sqrt{2}}|\up\up +\down\down\ra\\
|e_2\ra={i\over \sqrt{2}}|\up\up -\down\down\ra\\
 |e_3\ra={i\over \sqrt{2}}|\up\down +\down\up\ra\\
  |e_4\ra={1\over \sqrt{2}}|\up\down -\down\up\ra.
  \eea
Expanding in this complete set of functions one has
\bea  |\ph\ra=\sum_{j=1}^4\a_j|e_j\ra.\eea
The reduced density matrix is defined by taking the trace of the operator $|\ph\ra\la\ph|$ of either of the two particles.
The entanglement, $E$, of $ |\ph\ra$ can then be computed as the  von Neumann entropy of the reduced density matrix of either of the two particles. Ref.~\cite{Bennett:1996gf} found that the entanglement of $|\ph\ra$ can be expressed in terms of the entanglement entropy,
\bea H(x)\equiv -x \log_2(x) -(1-x)\log_2(1-x),\eea
which has a maximum of unity at $x=1/2$ and vanishes for $x=0,1$. 
One computes
 \bea C=\bigl|\sum_j\a_j^2\bigr|,
 \label{Cdef}
 \eea where one squares the complex numbers $\a_j$, and the result is that
\bea E(C)=H({1\over 2}\big(1+ \sqrt{1-C^2})\big).\label{Edef}\eea
The state of maximum entropy, with the density matrix proportional to the identity operator,  has $C=0$ and $E(C)=H(1)=0$-- no entanglement.  On the other hand, taking $\ph$ to be one of the Bell states gives $C=1$ and $E(C)=H(1/2)=1$, the maximum entanglement.


 BKKS   defined the entanglement
power of the $S$-matrix in a two-particle spin
space~\cite{Zanardi:2001zza} by the action of the $S$-matrix on an
incoming two-particle tensor product state with randomly-oriented
spins, 
\bea
|\psi_\text{in}\rangle = \hat R(\Omega_1)
|\hspace{-0.3em}\uparrow \rangle_1 \otimes \hat R(\Omega_2)
|\hspace{-0.3em}\uparrow \rangle_2 ,
\label{BKKS}\eea where $ \hat R(\Omega_j)$ is the
rotation operator acting in the $j^{\rm th}$ spin-${1\over 2}$ space.
This initial state is achieved in experiments by having a polarized beam impinge on a polarized target with all possible orientations available. No present experimental set up  can achieve that situation.
The two-particle density matrix of the final state is then  
$\hat\rho_{12} = |\psi_\text{out}\rangle\langle \psi_\text{out}| $ with $|\psi_\text{out} \rangle =
\hat {\bf S}|\psi_\text{in}\rangle$. The entanglement power, ${\mathcal E}$, of the $S$-matrix, $\hat {\bf
  S}$, is then~\cite{Beane:2018oxh}
\begin{equation}
{\mathcal E}({\hat {\bf S}})  = 1- \int {d\Omega_1\over 4\pi} \ {d\Omega_2\over 4\pi}\  {\rm Tr}_1\left[ \ \hat\rho_1^2 \  \right],
\ \ \ 
\label{eq:epgen}
\end{equation}
where $\hat\rho_1 = {\rm Tr}_2\left[\ \hat\rho_{12}\ \right]$ is the
reduced density matrix for particle 1 that acts in a space of dimension $d=2$.  

At sufficiently low energies the action of the $S$-matrix changes the amplitudes of    the two states with total spin $S=0,1$,
in the
$\si$ and $\siii$ channels.
%
BKKS studied 
the spin-space entanglement of two distinguishable particles, the
proton (1) and neutron (2).
Neglecting the  tensor-force-induced mixing of the $\siii$ channel with the $\diii$ channel,  
the $S$-matrix was expressed in terms of the $\si$ and $\siii$ 
phase shifts $\delta_{0,1},$ 
%
%
the entanglement power of $ \hat {\bf S}$ was calculated to be
%
${\mathcal E}({\hat {\bf S}})=
{1\over 6}\ \sin^2\left(2(\delta_1-\delta_0)\right),
$
which vanishes when $\delta_1-\delta_0= m {\pi\over 2}$ for any
integer $m$.  
But ${\mathcal E}({\hat {\bf S}})$ is maximal when the difference in phase shifts is $\pi/4$.
The triplet phase shift at 0 energy is $\pi$ because of the presence of the deuteron bound state and decreases with increasing energy. The singlet phase shift vanishes at 0 energy and increases as the energy increases from zero for low energies.  Thus the difference must pass through $\pi/4$ Indeed, using the phase shifts of \cite{PhysRevC.48.792} one finds that the difference passes through  $\pi/4$.
 at a lab energy of around 8.7 MeV  and  ${\mathcal E}({\hat {\bf S}})$ is maximized at that energy.

The quantity ${\mathcal E}({\hat {\bf S}})$  was evaluated as
a function of the center-of-mass nucleon momentum, $p$, (up to a lab energy of 350 MeV).
BKKS focused on values of $p$ between about 250 and 350 MeV/c,  finding that the ${\mathcal E}({\hat {\bf S}})
\approx 0.05$ and thus
  suppressed. However, the maximum value of ${\mathcal E}({\hat {\bf S}})$ is only 1/6 so that  ${\mathcal E}/ {\mathcal E}_{\rm max}\approx 0.3$, which is not very small. Moreover at such energies  all of the measured phase shifts are needed to describe scattering. 

 Furthermore, there is a problem with using \eq{eq:epgen} to determine entanglement. Suppose the density matrix is that of maximum entropy, $\r_{\rm max}$. Then
$ 
{\rm Tr_2}\r_{\rm max}={\hat I_1\over 2},
 $ 
 where $\hat I_1$ is the identity operator of the subspace of particle 1.
 On the other hand, defining $\r_i\equiv |e_i\ra\la e_i|$ (for any of $i$ between one and four) and taking ${\rm Tr}_2$ also yields  
${\rm Tr}_2\r_i={\hat I_1\over 2}
, $ 
which is the same as that of the state of maximum entropy and zero entanglement. The use of either $\r_{\rm max}$ or $\r_i$ in \eq{eq:epgen} would yield the same value, namely ${\mathcal E}={1/2}$.

Here I present an alternative analysis using the  precise measurement of entanglement power of Ref.~\cite{Bennett:1996gf}. This is done  by starting with an initial  pure state of 0 entanglement:
 \bea |\ph_i\ra= |\up\down\ra= -i|e_3\ra+|e_4\ra .\eea 
 Here $C=0$ and $H=0$  from \eq{Cdef} and \eq{Edef}.
 The action of scattering produces a normalized density matrix of the form 
 \bea \r_f={M|\ph\ra\la \ph| M^\dagger\over {\rm Tr}[|\ph\ra\la \ph|M^\dagger]}
\eea 
where 
$M(\bfp_f,\bfp_i)$ is the 
  neutron-proton  scattering operator  acting in the two-nucleon spin space. 
Use of invariance principles  (parity, time reversal and isospin)~\cite{PhysRev.85.947} shows there are five independent amplitudes needed to capture the scattering amplitude.  In particular~\cite{HoshizakiApp},
\bea &M(\bfp_f,\bfp_i) = a +c\,( \hat  {\bm \sigma}_1 +  \hat  {\bm \sigma}_2)\cdot \hat {\bf n}+m\, \hat{\bm\s}_1\cdot  \hat {\bf n}\hat{\bm\s}_2\cdot  \hat {\bf n}\nonumber\\&+g\left[\hat  {\bm \sigma}_1\cdot\hat{\bf P}\hat  {\bm \sigma}_2\cdot\hat{\bf P}+\hat  {\bm \sigma}_1\cdot\hat{\bf K}\hat  {\bm \sigma}_2\cdot\hat{\bf K}\right]\nonumber\\&
+h\left[\hat  {\bm \sigma}_1\cdot\hat{\bf P}\hat  {\bm \sigma}_2\cdot\hat{\bf P}-\hat  {\bm \sigma}_1\cdot\hat{\bf K}\hat  {\bm \sigma}_2\cdot\hat{\bf K}\right].
\label{fullm}
\eea
The results presented here use the amplitudes from the   NN online website: https://nn-online.org that
are computed from the measured phase shifts of Ref.~\cite{PhysRevC.48.792}.

The first result, for  lab kinetic energy of 1 MeV is shown in Fig.~1 finds that entanglement is maximized at all scattering angles.
\begin{figure}[h]
\label{1MeV}
  \centering
   \includegraphics[width=0.25\textwidth]{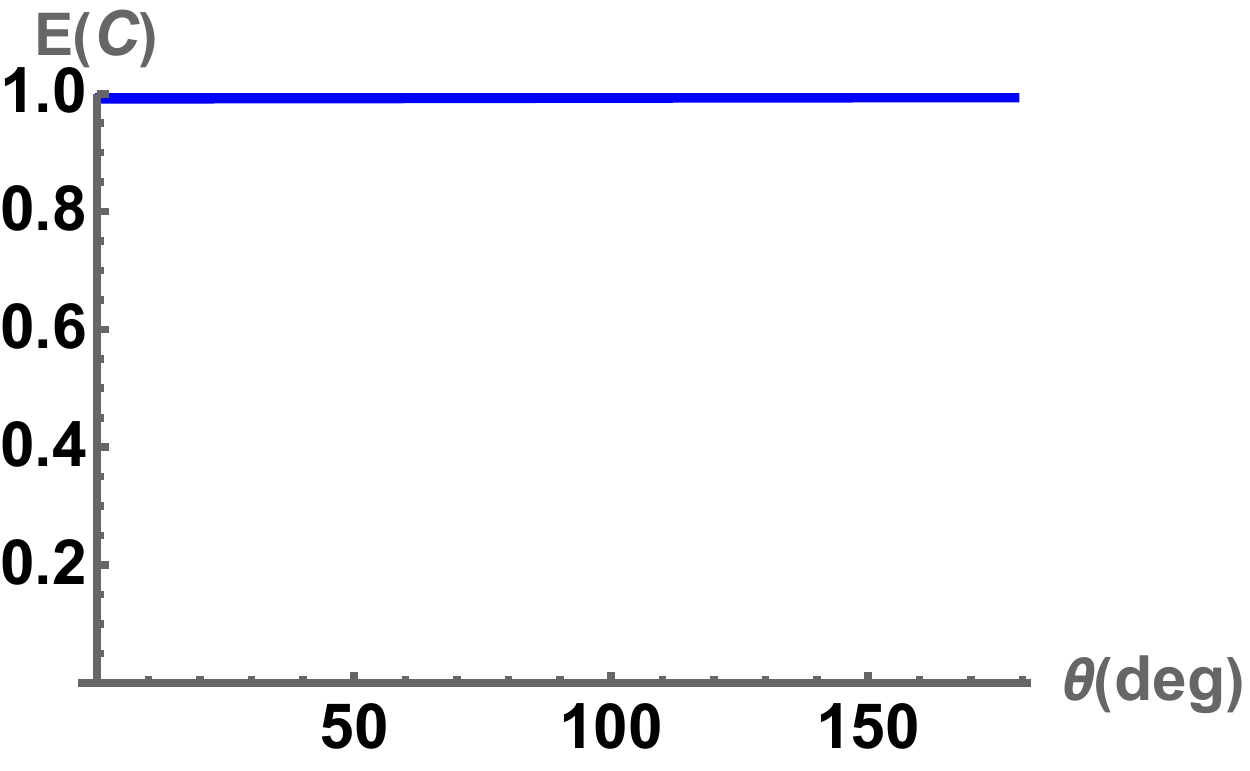}
  \caption{ Entanglement at 1 MeV 
   Computations use the  phase shifts of Ref.~\cite{PhysRevC.48.792}.  The state is $M|\up\down\ra$
 }
 \end{figure}
This result  can be understood by assuming that only s-waves contribute, approximately true at 1 MeV.
  In that case, $c=0$,  $h=0$ and $m=g$, which means that $M$ can be expressed as $M_L\equiv M_0+M_1\bfsigma_1\cdot\bfsigma_2$. Then using Eq(13.2) of Ref.~\cite{HoshizakiApp} the operator $M$ can be expressed in terms of  Bell states as:
\bea M_L|\ph_i\ra= {-i\over \sqrt{2}}
 (a+m)|e_3 \ra +{1\over \sqrt{2}} (a-3m) |e_4\ra.
 \eea
 At very low energies $a+m\propto e^{i \d_1}\sin\d_1,$ and $a-3m \propto e^{i \d_0}\sin\d_0$.
 Then a direct computation leads to the result 
 \bea 1-C^2={4\sin^2\d_1\sin^2\d_0\cos^2(\d_1-\d_0)\over \sin^2\d_0+\sin^2\d_1},
 \eea
 so that $C=0$   and $H=1$ when the phase shifts differ by $\pi/2$.   The triplet phase shift is $\pi$ at 0 energy because of the deuteron bound state in that channel. It drops rapidly with increasing lab energy. The singlet phase shifts vanishes at 0 energy and increases rapidly with energy. Thus a phase shift difference of $\pi/2$ is inevitable and  occurs at about 1 MeV as shown in 
 Fig.~2.

\begin{figure}[h]
  \centering
     \includegraphics[width=0.3\textwidth]{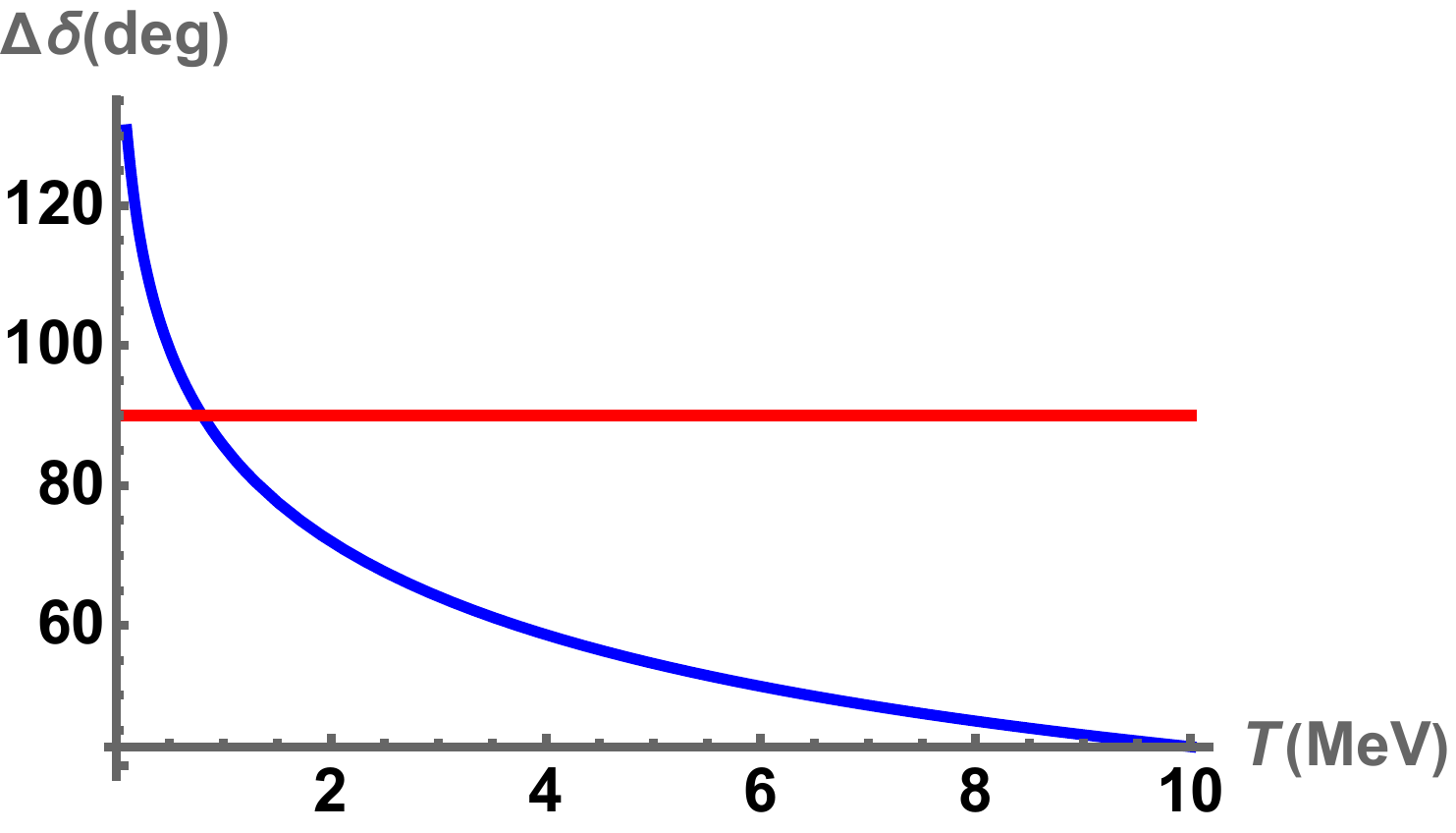}
  \caption{ Phase shift difference $\D\d\equiv\d_1-\d_0$. The phase shifts of Ref.~\cite{PhysRevC.48.792} are used.
 }
  \label{Phase}
 \end{figure}
 
 The result of Fig.~1 can be interpreted in terms of Wigner SU(4) symmetry~\cite{Wigner:1936dx,PhysRevLett.82.2060}.
 A nuclear Hamiltonian consistent with SU(4) symmetry obeys 
 \bea \large[H,\sum_i\vec \tau_i\Large]= \large[H,\sum_i\vec \s_i\Large]= \large[H,\sum_i\vec \tau_i\vec\s_i\Large]=0.
 \eea
 At sufficiently low energies for which the scattering is described using  s-wave phase shifts as the matrix 
 $M_L$, and the two-nucleon potential can be expressed in the same way~\cite{Ericson:1988gk}.
 In that case, the Hamiltonian satisfies SU(4) symmetry and that symmetry is consistent with maximum entanglement. However at higher energies, all of the terms of \eq{fullm} enter into the two-nucleon potential and SU(4) symmetry is broken. In that case one may expect to observe a different set of results for $E(C)$. 
 
 The results for lab kinetic energies up to 50 MeV  are shown in Fig.~3.  Observe that the angular dependence varies rapidly as the lab kinetic energy is increased from 1 to 50 MeV. This is due to the rapid dependence of the s-wave phase shifts on energy and the increasing importance of d-, p- and f- waves.
 
 \begin{figure}[h]
  \centering
     \includegraphics[width=0.4\textwidth]{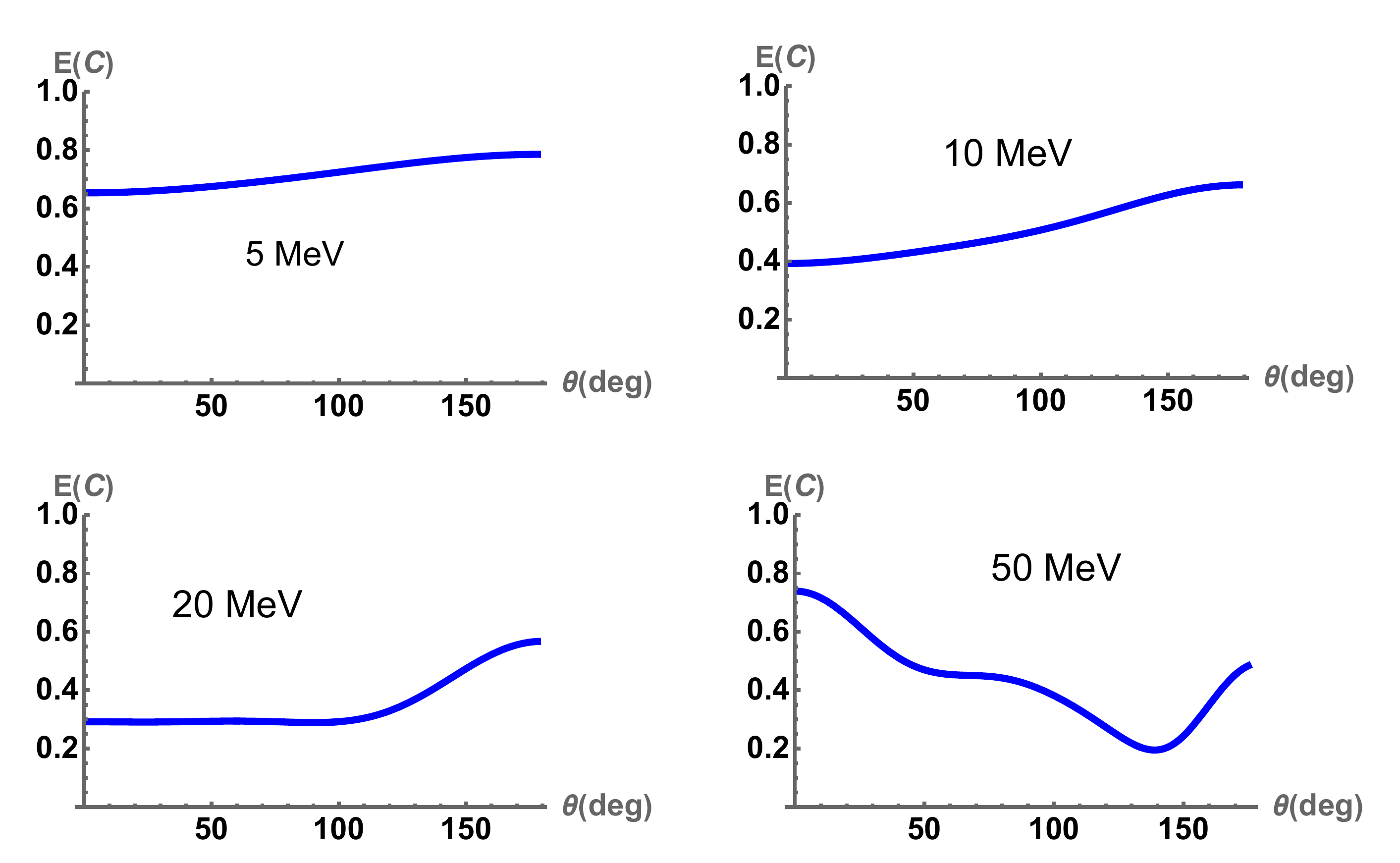}
  \caption{ $E(C)$ of \eq{Edef} for several lab kinetic energies as a function of center of momentum angles.  The state is $M|\up\down\ra$.
 }
  \label{LowE}
 \end{figure}

 \begin{figure}[h]
  \centering
     \includegraphics[width=0.4\textwidth]{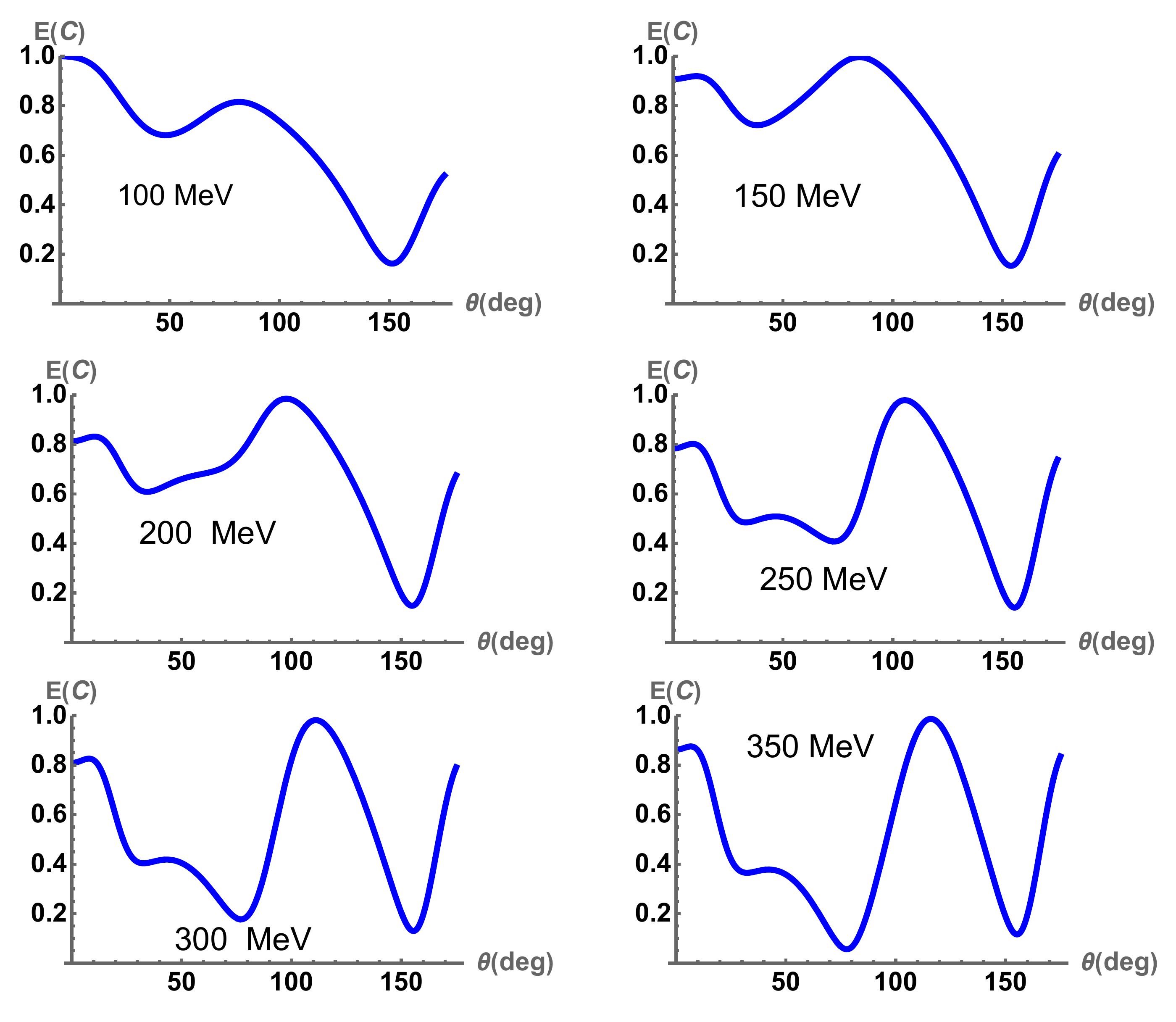}
  \caption{ $E(C)$ of \eq{Edef} for several lab kinetic energies as a function of center of momentum angles. The state is  $M|\up\down\ra.$
 }
  \label{HighE}
 \end{figure}

 The results for lab kinetic energies between   100  and 350 MeV  are shown in Fig.~4.  Observe the persistent prominent peak at around 90$^\circ$. It has long been known~\cite{Jastrow:1951vyc} that one-pion exchange is important for these energies, Forward-angle charge exchange allows n-p scattering to peak at backward angles and thus provide a signature.  The salient feature of one-pion exchange is the tensor force that is responsible for the binding of the deuteron.
 
 I compute the entanglement effect of the tensor operator 
 \bea& S_{12}=3 \bfsigma_1\cdot\hat\bfK\bfsigma_2\cdot\hat\bfK-\bfsigma_1\cdot\bfsigma_2
  \eea on the state $|\up\down\ra$. It is useful to use the Hoshizaki coordinate system:
  $\hat {\bf P}=(\sin\theta/2,0,\cos\theta/2),\,\hat\bfn =(0,1,0),\,\hat\bfK=(\cos\theta/2,0,-\sin\th/2)$. The operator $S_{12}$ acts only on triplet states, so the state $|\up\down\ra $ is projected to the triplet state with magnetic quantum number 0, $|\chi_0\ra/\sqrt{2}=-i |e_3\ra/\sqrt{2}$.
  Then a calculation yields
  \bea S_{12}|\up\down\ra= {i\over\sqrt{2}}[(3\cos\th -1)|e_3\ra+3\sin\th|e_2\ra]\eea
  a completely entangled state that has $E(C)=1$.  Thus it is reasonable to suggest that the large values of $E(C)$ seen in Fig.~4  for non-zero values of $\th$ result from the tensor force in combination with the other components of the nuclear force.

 One could also start with the state $|\up\up\ra={1\over\sqrt{2}} (|e_1\ra-i|e_2\ra$. This is also a direct product state with $C=0$ and 0 entanglement entropy.
 In the s-wave limit   the action of the scattering operator leaves the state invariant because this state is a spin eigenstate. 
 The computed values of $E(C)$ vanish for lab kinetic energies below  about   50 MeV. For higher energies there is an interesting angular dependence that displays significant entanglement.  The results are shown in Fig.~5.  
 \begin{figure}[h]
  \centering
     \includegraphics[width=0.4\textwidth]{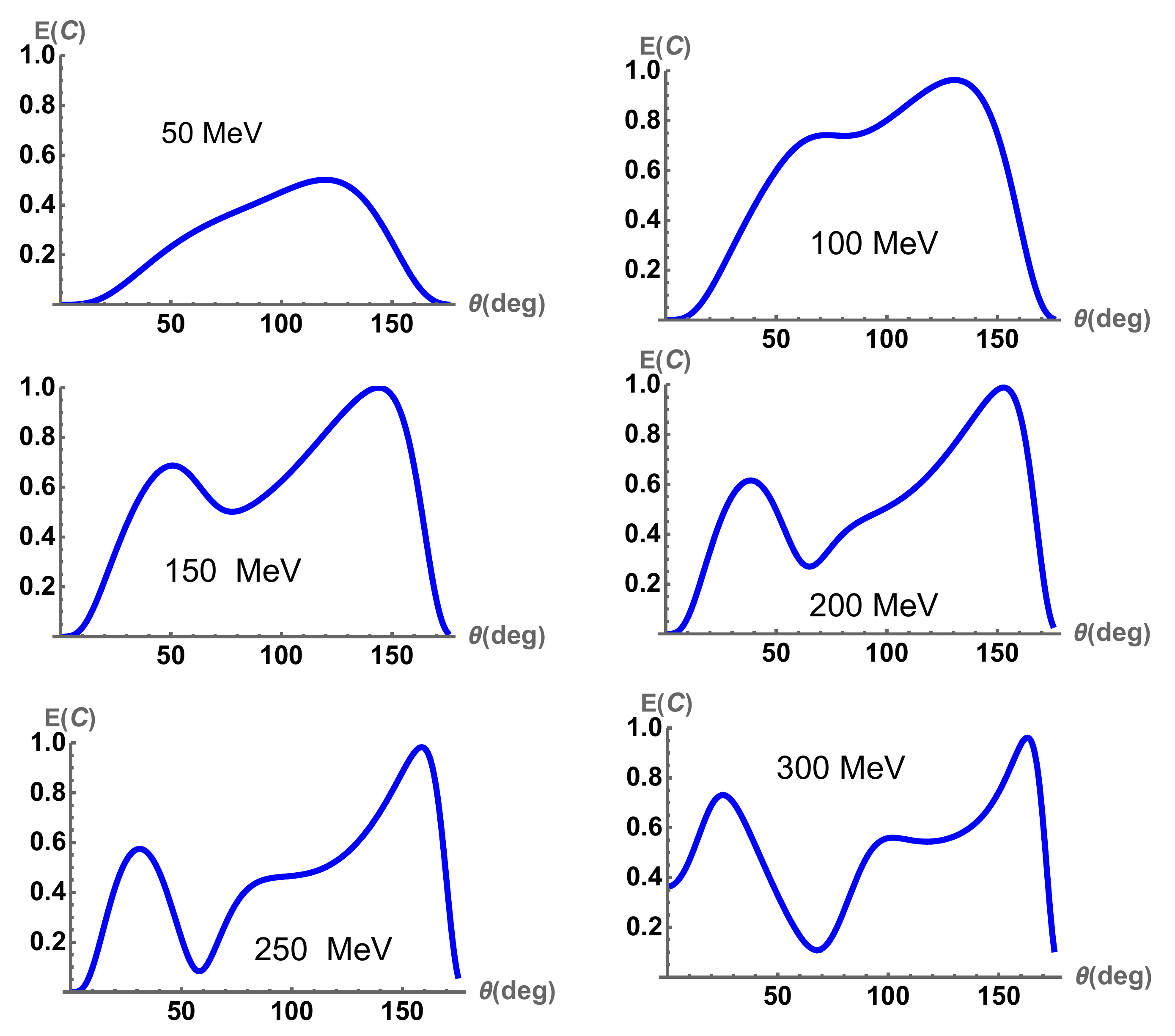}
  \caption{ $E(C)$ of \eq{Edef} for several lab kinetic energies as a function of center of momentum angles. The state is $M|\up\up\ra$.
 }
  \label{upup}
 \end{figure}

 Observe that the entanglement is generally large. Once again the effects of the tensor force are prominent because
 \bea&
 S_{12}|\up\up\ra\nonumber\\&={1\over\sqrt{2}}(2|e_1\ra+i(3\cos\th+1)|e_2\ra-3 i\sin\th|e_3\ra)
 ,
 \label{t}\eea
 a state that has $C={3(1+\cos\th)\over 7+3\cos\th} $. The related entanglement   is   shown in 
 Fig~6. The tensor effect of \eq{t} does not fully account for the results of Fig.~5, but does provide a substantial contribution.

 \begin{figure}[h]
  \centering
     \includegraphics[width=0.34\textwidth]{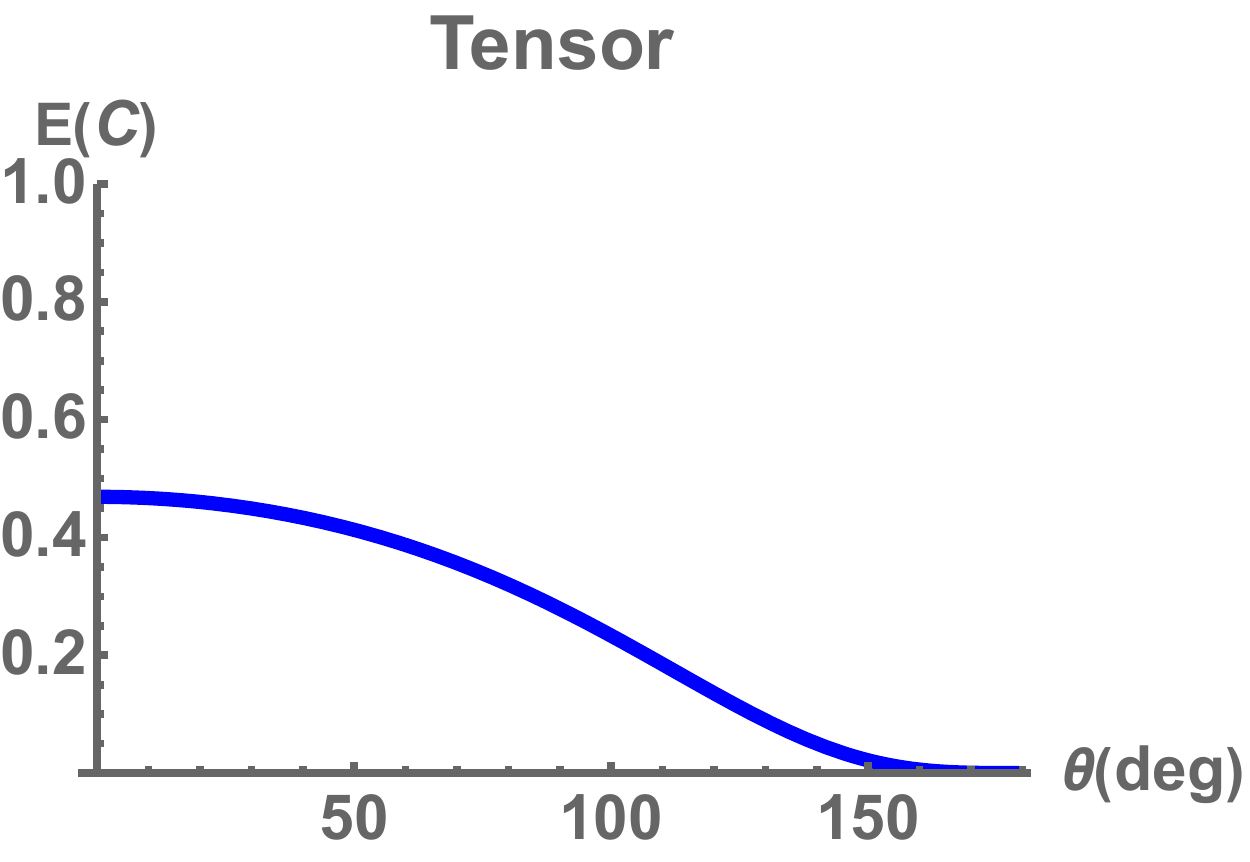}
  \caption{ Tensor contribution of \eq{t} to $E(C)$ of \eq{Edef}.
 }
  \label{upup}
 \end{figure} 

A summary is in order at this point. Entanglement is computed here using a technique~\cite{Bennett:1996gf}  that literally counts the number of entangled pairs produced by the neutron-proton interaction. Simply taking the trace  of the two-particle density matrix on particle 2 to obtain a one-body density matrix and computing the resultant entropy does not yield the entanglement entropy because
very completely entangled  and completely unentangled two-nucleon density matrices can yield the same on-particle density matrix.
 
Computations of  $E(C)$ of \eq{Edef} show  that entanglement is large for low-energy neutron-proton scattering. At such energies the nuclear potential satisfies Wigner SU(4) symmetry, so entanglement maximization is a sign of that symmetry. At higher energies the angular dependence
of entanglement is strong and is generally not suppressed.  The tensor force is shown to play a significant role in producing entanglement.

 
It is worth commenting on the role of symmetries in the entanglement properties of the two-nucleon interaction. The key feature used to obtain the present results is the limitation, caused by isospin, parity and time-reversal invariance, of the scattering operator to only five operators.
 Charge symmetry breaking, a violation of isospin invariance of high order in chiral power counting~\cite{vanKolck:1997fu},  leads to (class IV) operators of the form 
$(\t_1-\t_2)_z  (\bm \s_1-\bm \s_2)\cdot \bf\hat n$~\cite{Miller:1990iz}. Violations of parity would lead to operators of the form, for example, 
$ (\t_1-\t_2)_z\bm (\bm\s_1-\bm \s_2)\cdot (\bfp_i+\bfp_f)$~\cite{Gardner:2017xyl}  and time reversal violation would allow terms of the form
 $(\t_1-\t_2)_z  (\bm\s_1\times\bm \s_2)\cdot \bf\hat n$~\cite{Simonius:1975ve}.
  If the strength parameters governing all of these symmetry violations were of the size of other strong interaction terms, one would observe even greater entanglement.
Thus there are potentially deep connections between  entanglement and the fundamental symmetries of the Standard Model. 

This work was supported by the U. S. Department of Energy Office of Science, Office of Nuclear Physics under Award Number DE-FG02-97ER-41014. I thank Natalie Klco and Martin Savage for useful discussions.

 \bibliography{E}
 \end{document}